\documentclass[12pt]{article}
\usepackage{amsmath,amssymb}
\newcommand{\be}{\begin{equation}}
\newcommand{\ee}{\end{equation}}
\newcommand{\bea}{\begin{eqnarray}}
\newcommand{\eea}{\end{eqnarray}}

\def\P{Poincar\'e }

\begin{document}
\renewcommand {\thefootnote}{\fnsymbol{footnote}}
\vskip1cm
\begin{flushright}
\end{flushright}
\vskip1cm
\begin{center}
{\large\bf Twist Symmetry and Gauge Invariance}

\vskip .7cm {\bf{{M. Chaichian and A. Tureanu}}

{\it High Energy Physics Division, Department of Physical Sciences,
University of Helsinki\\
\ \ {and}\\
\ \ Helsinki Institute of Physics,\\ P.O. Box 64, FIN-00014
Helsinki, Finland}}
\end{center}
\vskip1cm
\begin{abstract}

By applying properly the concept of twist symmetry to the gauge
invariant theories, we arrive at the conclusion that previously
proposed in the literature noncommutative gauge theories, with the
use of $\star$-product, are the correct ones, which possess the
twisted Poincar\'e symmetry. At the same time, a recent approach to
twisted gauge transformations is in contradiction with the very
concept of gauge fields arising as a consequence of {\it local}
internal symmetry. Detailed explanations of this fact as well as the
origin of the discrepancy between the two approaches are presented.

\end{abstract}
\vskip1cm

\newpage

\section{Introduction}
The study of noncommutative quantum field theories (NC QFT) with
Heisenberg-like commutation relations \cite{SW} (for a review, see
\cite{Szabo}) has got a new impetus after it was realized that,
although they violate Lorentz invariance, they are however subject
to a Lorentz-invariant interpretation due to their twisted \P
symmetry \cite{CKT}.

The gauge invariance of noncommutative field theories has been
investigated for a long time, since the low-energy limit obtained
from string theory in the presence of a constant background field is
a noncommutative gauge theory, related to a commutative one by the
Seiberg-Witten map \cite{SW}. Noncommutative gauge theories have
been studied also in their own right, without the use of the
Seiberg-Witten map, and this study was initiated in \cite{Hayakawa},
by building NC QED. It was also understood that the use of the
$\star$-product imposes rather strict constraints on the
noncommutative gauge symmetry, among which was the fact that only NC
gauge $U(n)$ groups close (and not $SU(n)$). Moreover, there is a
no-go theorem \cite{nogo} stating that only certain representations
of the gauge group are allowed (fundamental, antifundamental and
adjoint) (see also \cite{terashima}) and the matter fields can be
charged under at most two gauge groups. Using these features of the
noncommutative gauge theories a noncommutative version of the
Standard Model was built \cite{NCSM,NCSM_ar}, with the gauge group
$U_\star(3)\times U_\star(2)\times U_\star(1)$, which solved the
problem of electric charge quantization in NC QED arrived at in
\cite{Hayakawa}.

Another noncommutative version of the Standard Model was built using
the Seiberg-Witten map, where the gauge invariance is defined with
respect to the infinitesimal local transformations of NC
$su(3)\times su(2)\times u(1)$ (by the use of the Seiberg-Witten
map, one can close these noncommutative gauge algebras
\cite{Wess_algebra} and some others as well \cite{Bonora}, however
not the corresponding gauge {\it groups}).

In this Letter we argue that the noncommutative gauge theories
constructed with the use of $\star$-product (i.e. $\star$-action of
the gauge algebra generators on the fields) remain the only correct
ones, possessing also twisted \P symmetry. We show that a recent
approach to twisted gauge transformations \cite{Vassilevich,Wess},
apparently allowing any gauge group to close, just as in the
commutative case, is in contradiction with the very idea of
introducing gauge fields when symmetry under {\it local}
transformation is required.

\section{Gauge transformations in NC field theory}
In all these approaches to noncommutative gauge theories
\cite{Hayakawa}-\cite{Bonora} and in the further developments based
on them, the essential aspect was that the gauge transformations of
the fields, whether infinitesimal or finite, were considered as
$\star$-gauge transformations. For example, in the case of the gauge
$U_\star(n)$ group, an arbitrary element of the group will be
\be U(x)=\exp_\star({i\alpha^a(x)T_a}) \ee
where $T_a$, $a=1,...,n^2$ are the generators of the $U(n)$ group,
with the algebra $[T_a,T_b]=if_{abc}T_c$, $\alpha^a(x)$,
$a=1,...,n^2$ are the gauge parameters and the $\star$-exponential
means
\be
\exp_\star({i\alpha^a(x)T_a})=1+i\alpha^a(x)T_a+\frac{i^2}{2!}\alpha^a(x)\star\alpha^b(x)T_aT_b+...\ee

Under the transformations of the $U_\star(n)$ gauge group, the
matter fields can be in the fundamental representation:
\be \psi(x)\rightarrow\psi'(x)=U(x)\star\psi(x) \ee
or antifundamental representation:
\be \chi(x)\rightarrow\chi'(x)=\chi(x)\star U^{-1}(x)\,  \ee
$U^{-1}(x)$ is the $\star$-inverse of $U(x)$, while the gauge fields
are, as they should, in the adjoint representation:
\be A_\mu(x)\rightarrow A_\mu'(x)=U(x)\star A_\mu(x)\star
U^{-1}(x)+i U(x)\star \partial_\mu U^{-1}(x)\ , \ee
where the matrix form of the gauge fields $A_\mu(x)=T_aA_\mu^a(x)$
is used. The covariant derivative, defined as usual
\be D_\mu=\partial_\mu-iA_\mu(x)\ ,\ee
but acting with $\star$:
\be D_\mu\psi(x)=\partial_\mu\psi(x)-iA_\mu(x)\star\psi(x)\ ,\ee
transforms appropriately, like the original fields
\be D_\mu\psi(x)\rightarrow D'_\mu\psi'(x)=U(x)\star
\left(D_\mu\psi(x)\right)\ ,\ee
while the field strength tensor $F_{\mu\nu}$
\be F_{\mu\nu}(x)=\partial_\mu A_\nu(x)-\partial_\nu A_\mu(x)
-i[A_\mu(x),A_\nu(x)]_\star \ee
can be easily shown to transform as
\be F_{\mu\nu}(x)\rightarrow F'_{\mu\nu}(x)=U(x)\star
F_{\mu\nu}(x)\star U^{-1}(x)\ ,\ee
such that the action of a noncommutative theory with fermionic
matter fields:
\be\label{action} S=\int d^4x \left[i\bar\psi\star \gamma^\mu
D_\mu\psi-m\bar\psi\star\psi-\frac{1}{4}\mbox{Tr}( F^{\mu\nu}\star
F_{\mu\nu})\right]\ee
is invariant under the noncommutative gauge group $U_\star(n)$. We
emphasize that the action (\ref{action}) is also twisted-\P
invariant.

\section{The concept of gauge invariance}
Recently, there have been attempts to approach gauge invariance of
noncommutative theories by using the mechanism of the twist
\cite{Vassilevich,Wess}, as explained in Section 4. We argue here
that the very approach of gauge transformation by the twist is in
contradiction with the gauge principle itself \cite{YM}. For this
purpose, we shall briefly review the introduction of the gauge field
in the usual commutative theory, following the classical paper of
Utiyama \cite{Utiyama} (for a pedagogical presentation, see
\cite{CN}).

Let us consider a Lagrangean density ${\cal
L}(\Phi_i(x),\partial_\mu \Phi_i(x))$, where $\Phi_i(x)$ are the
fields, and a Lie group of internal global transformations, $G$.
Under the infinitesimal transformations of the corresponding algebra
${\cal G}$, the fields and their derivatives transform as
\bea \Phi_i(x)\rightarrow \Phi_i'(x)=\Phi_i(x)+\delta \Phi_i(x),\ \ \ \ \ \delta \Phi_i(x)= iT_{ij}^a\alpha_a \Phi_j(x),\label{field_global}\\
\partial_\mu \Phi_i(x)\rightarrow \partial_\mu \Phi'_i(x)=\partial_\mu \Phi_i(x)+\delta (\partial_\mu \Phi_i(x)),\ \ \ \ \
\delta(\partial_\mu \Phi_i(x))= iT_{ij}^a\alpha_a
\partial_\mu \Phi_j(x)\label{deriv_global} ,\eea
where $T_{ij}^a$ are the generators of the group in component form
and $\alpha_a$ are the {\it global} parameters of the group. In
terms of group representations, (\ref{field_global}) and
(\ref{deriv_global}) show that, if a field $\Phi_i(x)$ is a
representation of the global Lie algebra ${\cal G}$, then its first
derivative with respect to space time $\partial_\mu \Phi_i(x)$ (and
actually its derivatives of any order) is also a representation of
${\cal G}$.

The invariance of the Lagrangean under the transformations of the
algebra ${\cal G}$ is expressed by the condition:
\be\frac{\partial{\cal L}}{\partial \Phi_i}\delta
\Phi_i+\frac{\partial{\cal L}}{\partial (\partial_\mu \Phi_i)}\delta
(\partial_\mu \Phi_i)=0\ ,\ee
which, upon taking into account (\ref{field_global}) and
(\ref{deriv_global}), becomes:
\be\frac{\partial{\cal L}}{\partial \Phi_i}T_{ij}^a\alpha_a
\Phi_i+\frac{\partial{\cal L}}{\partial (\partial_\mu
\Phi_i)}T_{ij}^a\alpha_a \partial_\mu \Phi_i=0\ .\ee

If now we make the transformations local by taking the infinitesimal
parameters dependent on coordinates, $\alpha_a(x)$, the
transformation of the fields will be of the same form as
(\ref{field_global})
\be\delta_\alpha \Phi_i(x)= iT_{ij}^a\alpha_a(x) \Phi_j(x)\ ,\ee
however, the variation of the derivatives, taking into account that
$\delta$ and $\partial_\mu$ act in different spaces and therefore
they commute, will read:
\be\label{gauge_deriv}\delta_\alpha(\partial_\mu \Phi_i(x))=
iT_{ij}^a\alpha_a(x)
\partial_\mu \Phi_j(x)+iT_{ij}^a\Phi_j(x)
\partial_\mu \alpha_a(x)\ .\ee
In other words, if the transformation is local (gauge), the
derivatives $\partial_\mu \Phi_i(x)$ are not representations of
the gauge algebra. This is the essential point of gauge
transformations. Consequently, the variation of the Lagrangean
density will be nonzero
\be\label{var_lagr}\delta {\cal L}=\frac{\partial{\cal L}}{\partial
(\partial_\mu \Phi_i)}iT_{ij}^a \Phi_i(x)\partial_\mu\alpha_a(x)\neq
0\ .\ee
Therefore, in order to achieve the invariance under the gauge
transformations, compensating (gauge) fields need to be introduced
into the Lagrangean, whose transformations would annihilate
(\ref{var_lagr}). The gauge fields, transforming as
\be \delta_\alpha A^a_\mu(x)=f_{abc}\alpha_b(x)
A_\mu^c(x)+\partial_\mu\alpha^a(x)\ee
enter the Lagrangean in the combination
\be D_\mu \Phi_i(x)=\partial_\mu
\Phi_i(x)-iT_{ij}^a\Phi_j(x)A_\mu^a(x)\ ,\ee
which is the covariant derivative, transforming like the original
field (and therefore being a representation of the gauge algebra),
i.e.
\be\delta_\alpha(D_\mu \Phi_i(x))=iT_{ij}^a\alpha_a(x)(D_\mu
\Phi_j(x))\ .\ee

The purpose of this review was to show that the usual derivatives of
the fields are not representations of the gauge algebra and this is
essential for the introduction of the gauge fields.

\section{Twist approach to noncommutative gauge transformations}

Twisted \P symmetry of noncommutative QFT is important because of
the Lorentz-invariant interpretation (in the sense of one-particle
wave-functions) which it provides for a theory which effectively
breaks Lorentz symmetry \cite{CKT}. This interpretation is
exclusively due to the content of the representation theory in the
noncommutative case, which is the same as in the commutative case.
Essential for twisting the coproduct of the Lorentz generators with
the Abelian twist
\be\label{twist}{\cal
F}=\exp\left({\frac{i}{2}\theta^{\mu\nu}P_\mu\otimes
P_\nu}\right)\ee
is the fact that, once a field is a representation of the Lorentz
generators $M_{\mu\nu}$, any derivative of any order of the field is
still a representation of $M_{\mu\nu}$.

The idea of the works \cite{Vassilevich,Wess} was that, since the
gauge generators, defined as
\be \alpha(x)=\alpha^a(x)T_a \ee
do not commute with the generators of the \P algebra (in particular,
with the momentum generator, $P_\mu$), one could extend the \P
algebra by semidirect product with the gauge generators and apply
the twist (\ref{twist}) also to the coproduct of the gauge
generators
\be\label{coproduct}\Delta_0(\delta_\alpha(x))=\delta_\alpha(x)\otimes1+1\otimes\delta_\alpha(x)\rightarrow\Delta_t(\delta_\alpha(x))={\cal
F}\Delta_0(\delta_\alpha(x)){\cal F}^{-1}\ .\ee
We recall that twisting the coproduct of the generators of the \P
algebra requires a consistent deformation of the product of fields:
\be\label{star} m\circ(\phi\otimes\psi)=\phi\psi\rightarrow
m_\star\circ(\phi\otimes\psi)=m\circ{\cal
F}^{-1}(\phi\otimes\psi)\equiv \phi\star\psi\ .\ee

Since the procedure of the twist states clearly that, for
consistency, $\star$-products appear only in the algebra of the
representations of the \P algebra (i.e., the usual product of fields
is replaced by $\star$-product, as in (\ref{star})), while the
generators act on the fields as usual, it was natural to take as
infinitesimal gauge transformation of the individual fields the
usual form (without $\star$-product):
\bea \delta_\alpha \Phi(x)&=&i\,\alpha(x)\Phi(x)\,,\cr
\delta_\alpha \Phi^\dagger(x)&=&-i\,\Phi^\dagger(x)\alpha(x)\
.\eea
However, the variation of a term of the Lagrangean written as a
$\star$-product of fields under the gauge transformation, reads:
\bea\label{gauge_VW} \delta_\alpha (\Phi^{\dagger}(x)\star
\Phi(x))&=&m_\star\circ\Delta_t(\alpha(x))(\Phi^\dagger(x)\otimes
\Phi(x))\cr&=&m\circ{\cal F}^{-1}{\cal F}\Delta_0(\alpha(x)){\cal
F}^{-1}(\Phi^\dagger(x)\otimes
\Phi(x))\cr&=&m\circ\Delta_0(\alpha(x)){\cal
F}^{-1}(\Phi^\dagger(x)\otimes \Phi(x))\ .\eea
It is then claimed \cite{Vassilevich,Wess} that the result of the
above action is
\be\delta_\alpha (\Phi^{\dagger}(x)\star \Phi(x))=i\alpha^a
(x)[-(\Phi^{\dagger}(x)T_a)\star \Phi(x)+\Phi^{\dagger}(x)\star (T_a
\Phi(x))]\ .\ee
This claim is based on the fact that it is considered that the
derivatives of any order of the field $\Phi (x)$ are in the same
representation of the gauge algebra as the field itself, i.e.:
\be\label{deriv_repr}\delta_\alpha((-i)^nP_{\mu_{1}}...P_{\mu_{n}}\Phi(x))=\delta_\alpha(\partial_{\mu_{1}}...\partial_{\mu_{n}}\Phi(x))=\alpha(x)(\partial_{\mu_{1}}...\partial_{\mu_{n}}\Phi(x))\
,\ee
because only in this case we have, from (\ref{gauge_VW}),
\bea\label{expl}\delta_\alpha (\Phi^{\dagger}\star
\Phi)&=&m\circ\Delta_0(\alpha){\cal F}^{-1}(\Phi^\dagger(x)\otimes
\Phi(x))\cr&=&(\delta_\alpha\Phi^{\dagger})\Phi+\Phi^{\dagger}(\delta_\alpha\Phi)\cr
&+&\sum_{n=1}^{\infty}\frac{(-i)^n}{n!}\theta^{\mu_1\nu_1}...\theta^{\mu_n\nu_n}[(\delta_\alpha
P_{\mu_1}...P_{\mu_n}\Phi^\dagger)(P_{\nu_1}...P_{\nu_n}\Phi)+(P_{\mu_1}...P_{\mu_n}\Phi^\dagger)(\delta_\alpha
P_{\nu_1}...P_{\nu_n}\Phi)]\cr
&=&(\delta_\alpha\Phi^{\dagger})\Phi(x)+\Phi^{\dagger}(\delta_\alpha\Phi)\cr
&+&\sum_{n=1}^{\infty}\frac{i^n}{n!}\theta^{\mu_1\nu_1}...\theta^{\mu_n\nu_n}[-i\alpha^a
(\partial_{\mu_1}...\partial_{\mu_n}\Phi^\dagger
T_a)(\partial_{\nu_1}...\partial_{\nu_n}\Phi)+(\partial_{\mu_1}...\partial_{\mu_n}\Phi^\dagger)(i\alpha^a
\partial_{\nu_1}...\partial_{\nu_n}(T_a\Phi)]\cr
&=&i\alpha^a [-(\Phi^{\dagger}(x)T_a)\star
\Phi(x)+\Phi^{\dagger}(x)\star (T_a \Phi(x))]\ .\eea
It is obvious that, were (\ref{expl}) correct, it would immediately
follow that any algebra which closes in the commutative case, would
trivially close in this case as well, since the gauge parameters
$\alpha^a(x)$ are not affected by the $\star$-product.

It is easy to see that, using (\ref{deriv_repr}), one obtains as
well
\be\label{kinetic}\delta_\alpha (\Phi^{\dagger}\star
\partial_\mu\Phi)=i\alpha^a [-(\Phi^{\dagger}(x)T_a)\star
\partial_\mu\Phi(x)+\Phi^{\dagger}(x)\star (T_a
\partial_\mu\Phi(x))]\ ,\ee
which shows that, upon gauge transforming the kinetic terms of the
Lagrangeans, the latter remain invariant, without any need for the
introduction of the gauge fields.

We can now see that the essence of the gauge invariance is in
contradiction with the essence of the twist approach to gauge
transformations advocated in \cite{Vassilevich,Wess}: the twist
requires that the fields and their derivatives of any order be
representations of the gauge generator, as in (\ref{deriv_repr}),
contrary to (\ref{gauge_deriv}), the crucial point of the gauge
invariance machinery.

Actually, (\ref{deriv_repr}) and (\ref{expl}) are correct only if
the parameters $\alpha^a$ do not depend on coordinates, i.e. for
global internal transformations, which would also explain
(\ref{kinetic}), but in this case the whole twist approach is
redundant.

We can therefore conclude that the approach of
\cite{Vassilevich,Wess} is indeed leading to global internal
transformations of $\star$-products of fields and that gauge
transformations apriorically cannot be implemented by twisting the
gauge generators with the twist element (\ref{twist}), because they
do not satisfy the condition (\ref{deriv_repr}).

\section{Conclusions}

We have shown that gauge transformations of the action of NC QFT
(\ref{action}) cannot be introduced by twisting with (\ref{twist})
the coproduct of the usual gauge generators. Such a procedure, to be
consistent, would require that if a field is transformed in a
representation of the gauge algebra, then its derivatives of any
order also transform according to the representations of the gauge
algebra, what is in obvious contradiction with the very concept of
gauge transformations.

This leaves us with the only option of formulating noncommutative
gauge theories via the $\star$-product, as initiated in
\cite{Hayakawa} or by using the Seiberg-Witten map. The latter
theories have the twisted \P symmetry built in.

A full understanding of the concept of twist symmetry is
important. In this respect, we would like to mention that an
improper use of the twist in the quantization of noncommutative
field theories may also erroneously lead to a violation of the
spin-statistics relation and the Pauli exclusion principle, as it
has been clarified in \cite{ss}. The proper use of the concept of
twist, as outlined in the present paper, will be important also in
constructing a noncommutative version of the gravitational theory
\cite{CT_grav}.

\vskip 0.3cm {\bf{Acknowledgements}}

We are grateful to  Claus Montonen, Kazuhiko Nishijima and Peter
Pre\v{s}najder for illuminating discussions.

\end{document}